\documentclass[lettersize,journal]{IEEEtran}
\usepackage{amsmath,amsfonts}
\usepackage{booktabs}
\usepackage{multirow}
\usepackage{algorithmic}
\usepackage{algorithm}
\usepackage{array}
\usepackage[caption=false,font=normalsize,labelfont=sf,textfont=sf]{subfig}
\usepackage{textcomp}
\usepackage{stfloats}
\usepackage{url}
\usepackage{verbatim}
\usepackage{graphicx}
\usepackage{cite}

\usepackage{xcolor}
\begin{document}

\title{DC-Motion: Decoupling Structure and Details via Discrete-Continuous Tokens for Human Motion Generation}

\author{Hequan Wang,
        ~Xue'an Chen,
        ~Jiaxu Zhang,
        ~Zhengbo Zhang,
        ~Zhigang Tu,~\IEEEmembership{Senior Member,~IEEE}

\thanks{Hequan Wang and Zhigang Tu are with the State Key Laboratory of Information Engineering in Surveying, Mapping and Remote Sensing, Wuhan University, Wuhan 430072, China.}

\thanks{Xue’an Chen is with Geophysical Exploration Brigade of Hubei Provincial Geological Bureau, Wuhan 430058, China.}

\thanks{Zhengbo Zhang is with the Information Systems Technology and Design Pillar, Singapore University of Technology and Design, 487372, Singapore.}

\thanks{Corresponding author: Zhengbo Zhang (Email: zhangzb@whu.edu.cn).}
\thanks{Hequan Wang and Xue'an Chen contribute equally, they are co-first authors.}
}

\markboth{Journal of \LaTeX\ Class Files,~Vol.~14, No.~8, August~2026}%
{Shell \MakeLowercase{\textit{et al.}}: A Sample Article Using IEEEtran.cls for IEEE Journals}


\maketitle

\begin{abstract}
Text-to-motion generation requires modeling both global action structure and fine-grained motion dynamics from natural language. Existing methods usually rely on a single representation space: continuous diffusion models generate motion by progressively denoising a continuous latent representation, which preserves smooth dynamics but lacks an explicit interface for compositional temporal planning; while vector-quantized discrete models encode motion as sequences of codebook indices and leverage token-level sequence modeling for generation, which improves semantic controllability but compresses all motion details into finite codebooks, inevitably discarding fine-grained dynamics that cannot be well approximated by the nearest codebook entries. We argue that this limitation comes from a representation-level mismatch. Action intent, phase transition, and temporal layout are naturally discrete and compositional, whereas joint trajectories, contact timing, and velocity profiles are continuous and locally correlated. To address this mismatch, we propose DC-Motion, a discrete-continuous factorized framework for human motion generation. Instead of modeling motion as a purely continuous latent or a purely discrete token sequence, DC-Motion decomposes motion into discrete structural tokens for action layout and continuous residual latents for details that are not represented by the codebook. A text-conditioned structure generator predicts the discrete component through iterative masked prediction, and a residual diffusion model generates the continuous complement conditioned on the predicted structure. Unlike the residual vector quantization methods that approximate details with additional discrete indices, our DC-Motion keeps residual latent continuous. Experiments on HumanML3D and KIT-ML show DC-Motion achieves the best FID and R-Precision among strong diffusion-based and discrete-token baselines. 
\end{abstract}


\section{Introduction}
\IEEEPARstart{G}{enerating} human motion from text~\cite{cai2026coordinating, ahuja2019language2pose, zhang2024modular, kim2023flame, petrovich2022temos, lu2023humantomato, zhang2024motiondiffuse} is an active research area in computer vision and graphics, supporting applications in virtual reality~\cite{dang2026segmo}, digital human animation~\cite{gong2026diffusion, zhang2023skinned, zhang2024modular}, co-speech motion generation~\cite{zhang2025semtalk}, and robotics~\cite{gang2025strong}. Recent deep generative models approach this task mainly through two representation paradigms: continuous latent generation and discrete token generation. Continuous diffusion models~\cite{cai2026coordinating,ahn2018text2action, chen2023executing, zhang2023remodiffuse, yu2026causal} and latent flow-based models~\cite{dong2026motionflow} generate motion in continuous latent spaces and are particularly effective at producing smooth dynamics. In contrast, discrete token methods~\cite{zhang2023generating, jiang2023motiongpt, guo2024momask, li2026llamo, wang2026temporal} quantize motion into codebook indices and enable compact sequence modeling for text-conditioned generation.

Despite their progress, both paradigms use a single uniform representation space to jointly encode structurally diverse motion information, without distinguishing between high-level action structure and low-level motion dynamics. Human motion contains high-level structural factors, \textit{e.g.} action intent, temporal layout, and phase transitions, as well as low-level dynamic factors, \textit{e.g.} joint trajectories, contact timing, and velocity profiles~\cite{athanasiou2022teach, jiang2023motiongpt, taylor2006modeling}. Continuous latent models represent structural layout and fine dynamics within a single trajectory distribution, but they do not expose an intermediate structure for compositional temporal planning~\cite{chen2023executing, zhang2024motiondiffuse}. Purely discrete models take the opposite route by encoding both action structure and local motion details as finite codebook indices, which can discard dynamics that are poorly matched to the nearest codebook entries~\cite{zhang2023generating, jiang2023motiongpt, guo2024momask}. Recent discrete token methods such as MoMask~\cite{guo2024momask} introduce 
residual vector quantization to recover finer motion details through additional 
discrete codebook levels together with masked token prediction. However, 
representing residual details with additional discrete codebook indices 
inherently limits reconstruction fidelity, as any motion information not 
well-approximated by the nearest codebook entry is inevitably lost. s
\begin{figure*}[t]
    \centering
    \includegraphics[width=0.65\textwidth]{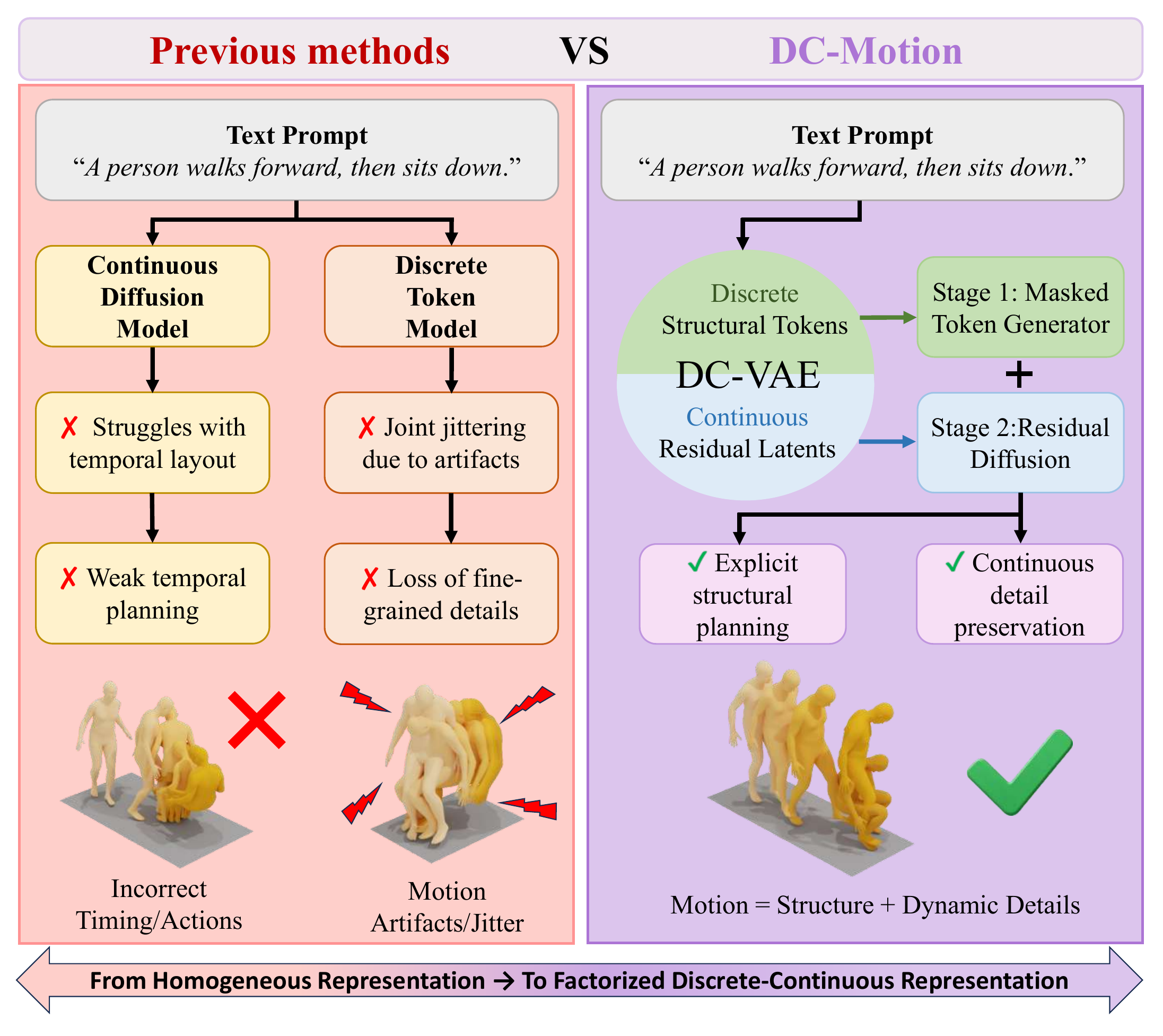}
    \caption{Overview of the characteristics of our DC-Motion. The proposed framework represents motion with discrete structural tokens and continuous residual latents, enabling text-conditioned structural generation followed by residual detail modeling. Darker colors indicate later time steps.}
    \label{fig:teaser}
\end{figure*}

Addressing this question calls for a representation that is simultaneously discrete enough to support structural abstraction and continuous enough to preserve fine-grained motion details. However, building such a hybrid discrete-continuous representation is non-trivial. On one hand, the discrete and continuous components must be complementary rather than redundant: the discrete part should capture structural layout, while the continuous part should model only the information left unexpressed by the discrete codebook. On the other hand, the generation process must be factorized accordingly, so that each generator is explicitly matched to its own component rather than independently modeling the full motion distribution. In this paper, we address this question with DC-Motion, a novel discrete-continuous factorized framework for text-to-motion generation. Specifically, we introduce a Discrete-Continuous Autoencoding Tokenizer (DC-VAE), which encodes motion into a continuous latent space, quantizes it into discrete structural tokens, and preserves the resulting quantization residual as a complementary continuous latent component. The discrete tokens provide a compact structural abstraction of motion, while the residual captures the remaining latent information not explicitly represented by the quantized prototypes.

Built upon this discrete-continuous representation, DC-Motion decomposes generation into two complementary stages: a text-conditioned masked token generator that predicts discrete structural tokens to capture global action layout, and a residual diffusion model that generates continuous residual latents to recover the fine-grained motion details left unexpressed by the discrete codebook. First, a text-conditioned structural token generator predicts discrete structural tokens through iterative masked prediction with bidirectional attention. Operating in the compact structural token space, this generator models the global action layout without generating fine-grained motion details. Second, a residual diffusion model generates continuous residual latents conditioned on both the text description and the predicted structural tokens. Since the residual contains only the information not represented by the discrete structural tokens, the diffusion model focuses exclusively on modeling the remaining motion details. Consequently, DC-Motion is not a simple combination of masked token generation and diffusion; rather, each generator is matched to a distinct component of the proposed representation.

We evaluate DC-Motion on HumanML3D and KIT Motion-Language using standard text-to-motion metrics, including Fréchet Inception Distance, R-Precision, Multi-Modal Distance, diversity, and multi-modality. On HumanML3D, DC-Motion achieves an FID of 0.041 and an R@1 of 0.528, improving over the strong residual-quantization baseline MoMask in both motion quality and text-motion alignment. On KIT-ML, DC-Motion obtains the lowest FID of 0.148 and the best R@1 of 0.442 among the compared methods. Ablation studies further show that the continuous residual branch, the discrete structural condition, and iterative masked structural prediction each contribute to the final performance.

The main contributions are summarized as follows:
\begin{itemize}
    \item We propose DC-Motion, a discrete-continuous representation framework that explicitly decouples structural abstraction from continuous residual detail modeling for text-to-motion generation.

    \item We introduce DC-VAE, a Discrete-Continuous Autoencoding Tokenizer, which quantizes motion into discrete structural tokens meanwhile retaining the quantization residual as a continuous latent component.
    
    \item We design a factorized generation process, in which a text-conditioned structural token generator predicts discrete motion structure and a residual diffusion model generates the continuous complement. Experiments on HumanML3D and KIT-ML show that this design improves both motion quality and text-motion alignment, achieving the best FID and R-Precision among the compared diffusion-based and discrete-token baselines.
    
\end{itemize}

\section{Related Work}

\subsection{Text-Driven Human Motion Synthesis}
The generation of human motion from natural language has witnessed significant progress. Beyond full-body human motion synthesis, related generative motion tasks have also been studied for hand and gesture motion~\cite{chen2025motion, zhou2025hand}, facial animation~\cite{cudeiro2019capture, fan2022faceformer}, and animal motion modeling~\cite{niewiadomski2025generative, zuffi20173d, zuffi2018lions}. Early approaches primarily relied on recurrent neural networks and generative adversarial networks~\cite{ahuja2019language2pose}. More recently, continuous diffusion models---such as MotionDiffuse~\cite{zhang2024motiondiffuse}, MDM~\cite{tevet2022human}, and MLD~\cite{chen2023executing}---have come to dominate the field by modeling motion in continuous spaces, which enables smooth motion dynamics. Diffusion-based motion models have also been explored for diverse human motion prediction, where DivDiff~\cite{yu2024divdiff} introduces skeletal constraints into the conditional diffusion process to improve prediction diversity and plausibility. However, these continuous models lack an explicit intermediate representation for compositional temporal planning, making it difficult to model motion structure in a structured and controllable manner.

In contrast, discrete token generation methods such as T2M-GPT~\cite{zhang2023generating} and MotionGPT~\cite{jiang2023motiongpt} map motion into codebook indices and leverage Transformer architectures to achieve stronger semantic alignment with text prompts. Scene-aware motion synthesis methods~\cite{gao2025jointly} further show that language commands and contextual intentions are important for generating semantically consistent 3D human motions. Subsequent work, including MMM~\cite{pinyoanuntapong2024mmm} and MoMask~\cite{guo2024momask}, advances discrete motion generation through masked token prediction and residual vector quantization, while Light-T2M~\cite{zeng2025light} investigates lightweight architectures for efficient text-to-motion synthesis. Nevertheless, purely discrete representations inherently compress both motion structure and fine-grained motion details into finite codebooks, discarding motion details that cannot be well approximated by the nearest codebook entries~\cite{zhang2023generating, jiang2023motiongpt, guo2024momask}. To overcome these limitations, DC-Motion introduces a discrete-continuous factorized representation that uses discrete structural tokens for action-level organization and continuous residual latents for fine-grained motion details.

\subsection{Motion Representation and Tokenization}
The choice of representation space is critical for modeling human motion. Prior studies on pose estimation and skeleton-based action recognition have shown that spatio-temporal dependencies and joint-bone relations provide important cues for human motion understanding~\cite{zhang2022mixste, tu2022joint}. Existing motion generation methods generally adopt either continuous latent spaces or discrete token spaces. Continuous autoencoder and latent diffusion methods, such as ACTOR~\cite{petrovich2021action} and MLD~\cite{chen2023executing}, preserve fine-grained motion details but require the model to capture action structure and local dynamics in a single continuous space, without an explicit mechanism to separate high-level temporal organization from low-level motion variations

Discrete token methods instead use vector-quantized autoencoders to represent motion as codebook indices, enabling compact sequence modeling~\cite{van2017neural, zhang2023generating}. Cross-modal quantization has also been explored for co-speech gesture generation~\cite{wang2024cross}. To improve reconstruction fidelity, MoMask~\cite{guo2024momask} and Mogo~\cite{fu2024mogo} further introduce residual vector quantization (RVQ) with multiple discrete codebook levels. However, RVQ remains purely discrete: residual details are still approximated by codebook indices, and information not captured by any codebook level is discarded. In contrast, DC-VAE quantizes motion into discrete structural tokens while retaining the quantization residual as a continuous latent, providing a continuous channel for fine-grained details beyond the discrete codebook.

\section{Method}

\subsection{Overview: Discrete-Continuous Factorization}
\label{sec:overview}

\begin{figure*}[t]
    \centering
    \includegraphics[width=\textwidth]{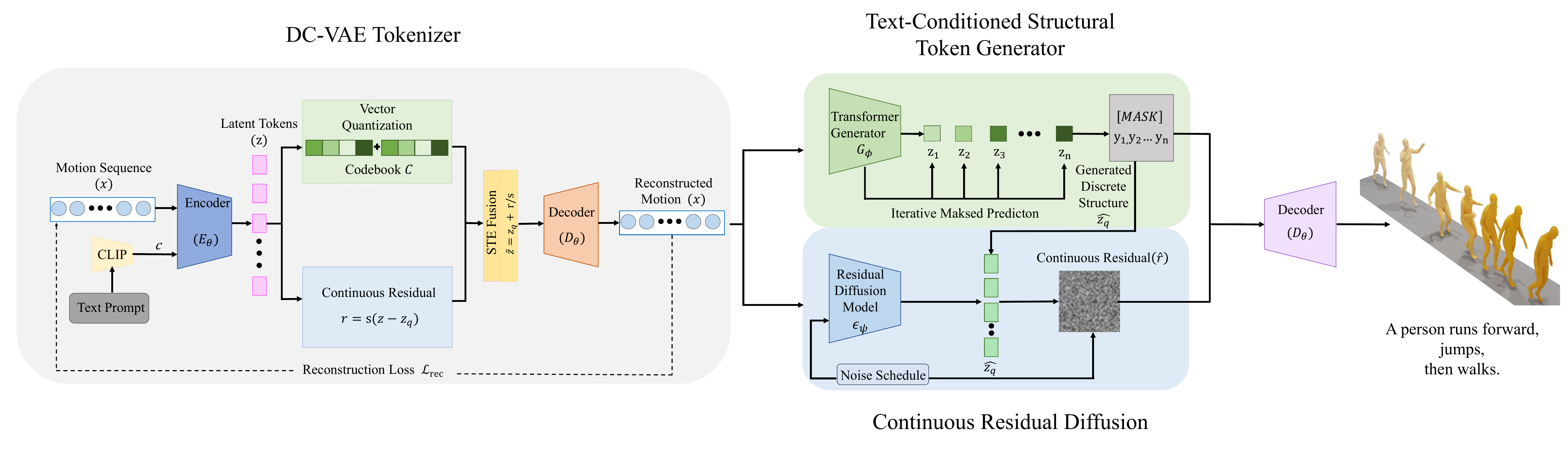}
    \caption{Overview of DC-Motion. The proposed framework first learns a DC-VAE tokenizer to decompose motion into discrete structural tokens and continuous residual latents. A text-conditioned structural token generator predicts the discrete structure through iterative masked prediction, while a residual diffusion model generates the continuous residual conditioned on both the text and the predicted structure. The final motion is decoded by combining the discrete prototype with the generated residual.}
    \label{fig:method}
\end{figure*}

Given a text description $c$, the goal of text-to-motion generation is to synthesize a human motion sequence $x_{1:T}=\{x_t\}_{t=1}^{T}\in\mathbb{R}^{T\times F}$ that is semantically aligned with $c$, where $T$ denotes the sequence length and $F$ denotes the per-frame feature dimension. Instead of modeling motion as a single continuous latent trajectory or a purely discrete token sequence, DC-Motion factorizes the motion latent into two complementary components: a discrete structural prototype and a continuous residual.

The key idea is to use vector quantization not as the final motion representation, but as a decomposition operator. Given a continuous latent token $z_i$, nearest-neighbor quantization selects a codebook entry $z_{q,i}$. The difference between the continuous latent and its quantized prototype is retained as a continuous residual. With a residual scaling factor $s$, the latent representation can be written as
\begin{equation}
    z = z_q + r/s,
\end{equation}
where $z_q$ denotes the discrete prototype and $r$ denotes the scaled continuous residual. This decomposition makes the two components complementary: $z_q$ represents the part of the latent captured by the codebook, while $r$ represents the remaining information not expressed by the selected codebook entry.

Based on this representation, DC-Motion uses a factorized generation process. A text-conditioned masked token generator predicts the discrete structural tokens from the text condition, and a residual diffusion model generates the continuous residual conditioned on both the text and the predicted structure. The two generators are therefore not independent modules combined post hoc; they model the two components produced by the same discrete-continuous tokenizer. The overall framework is illustrated in Fig.~\ref{fig:method}.

The following sections describe each component in detail: Sec.~\ref{sec:dcvae} introduces the DC-VAE tokenizer, Sec.~\ref{sec:maskgen} presents text-conditioned structural token generation, Sec.~\ref{sec:resdiff} describes conditional residual diffusion, and Sec.~\ref{sec:train_infer} summarizes the inference procedure.

\subsection{DC-VAE: Autoencoding Tokenizer with Continuous Residuals}
\label{sec:dcvae}

DC-VAE is a discrete-continuous autoencoding tokenizer that maps a motion sequence into paired discrete and continuous latent components. An encoder $E_\theta$ first maps the input motion $x$ into a sequence of continuous latent tokens:
\begin{equation}
z = E_\theta(x) \in \mathbb{R}^{N\times D},
\end{equation}
where $N$ is the number of latent tokens and $D$ is the dimensionality of each latent token.

We introduce a learnable codebook $\mathcal{C}=\{e_k\}_{k=1}^{K}$, where $e_k\in\mathbb{R}^{D}$. For each latent token $z_i$, vector quantization selects the nearest codebook entry:
\begin{equation}
y_i = \arg\min_{k\in\{1,\ldots,K\}}\|z_i-e_k\|_2^2,\quad z_{q,i}=e_{y_i}.
\end{equation}
This produces a discrete index sequence $y=\{y_i\}_{i=1}^{N}$ and a quantized latent sequence $z_q$.

Instead of discarding the quantization residual, DC-VAE explicitly retains it as a continuous latent component:
\begin{equation}
r = s(z-z_q),
\end{equation}
where $s$ is a scaling factor used to normalize the residual magnitude. The residual is the component of the continuous latent not represented by the selected codebook entry. It therefore provides a continuous channel for local motion variations that are difficult to encode with finite codebook indices.

During decoding, the continuous latent is reconstructed from the paired components:
\begin{equation}
\tilde{z}=z_q+r/s,\quad \hat{x}=D_\theta(\tilde{z}).
\end{equation}
The tokenizer is optimized with a reconstruction loss and a VQ-style commitment objective:
\begin{equation}
\mathcal{L}_{\mathrm{vae}}=\mathcal{L}_{\mathrm{rec}}(x,\hat{x})
+\lambda\|\mathrm{sg}[z]-z_q\|_2^2
+\beta\|z-\mathrm{sg}[z_q]\|_2^2.
\end{equation}
Here, $\mathrm{sg}[\cdot]$ denotes the stop-gradient operator. This objective trains the codebook to provide a compact structural abstraction while retaining the continuous residual for reconstruction.
\begin{figure}[t]
    \centering
    \includegraphics[width=1.0\linewidth]{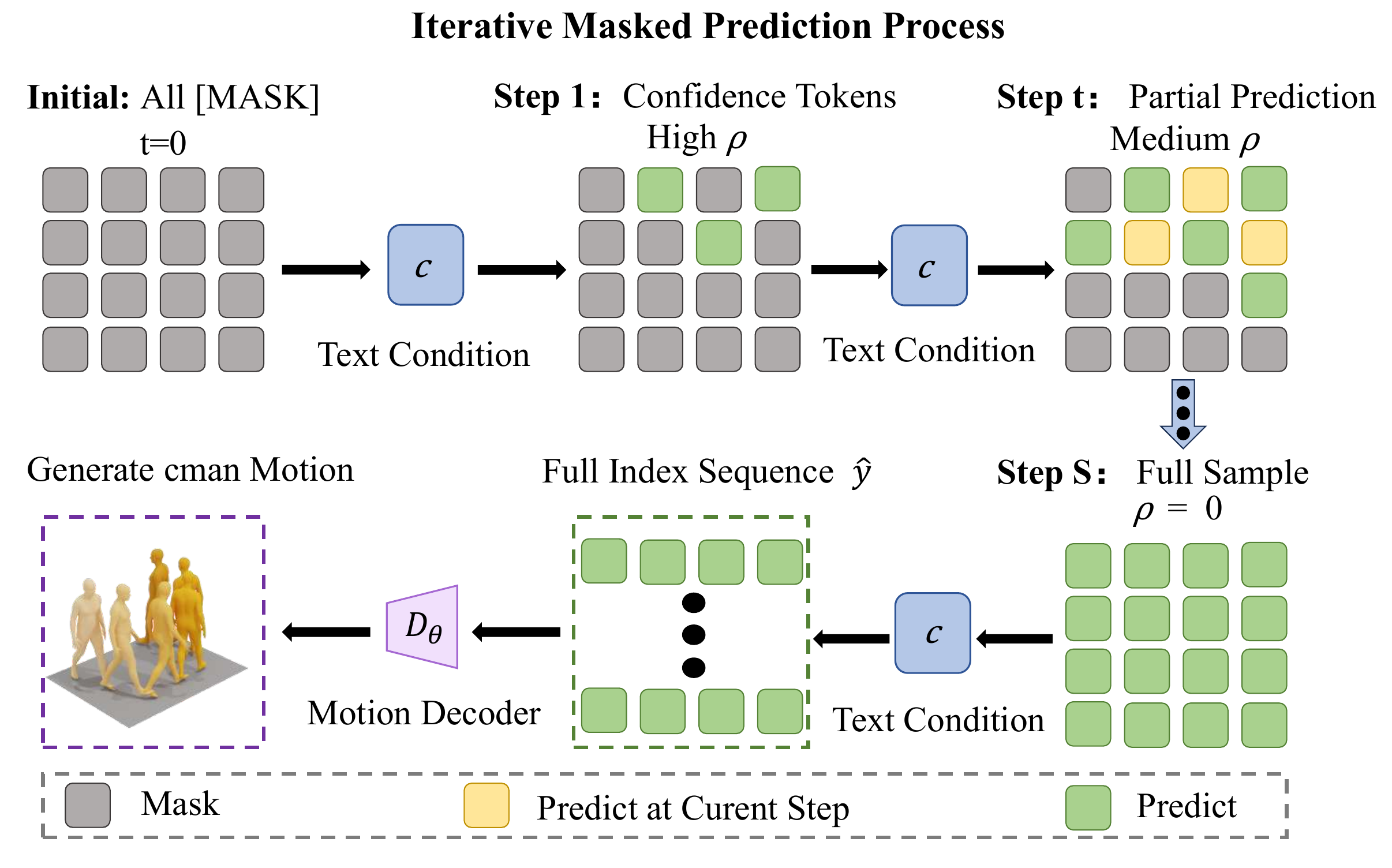}
    \caption{Iterative masked prediction process of the text-conditioned structural token generator $G_\phi$. Starting from a fully masked index sequence at $t=0$, $G_\phi$ predicts all masked positions in parallel conditioned on the text feature $c$. At each iteration, high-confidence predictions are fixed while the remaining low-confidence positions are re-masked for subsequent refinement. The masking ratio $\rho$ is progressively reduced from 1 to 0 according to a cosine annealing schedule over $S$ iterations, yielding the complete structural token sequence $\hat{y}$. The generated structural tokens are subsequently decoded by the shared motion decoder $D_\theta$ to produce the final motion sequence.}
    \label{fig:mask}
\end{figure}

\subsection{Text-Conditioned Structural Token Generation}
\label{sec:maskgen}

With the discrete index sequence $y \in \{1,\ldots,K\}^N$ produced by DC-VAE, we train a text-conditioned masked token generator $G_\phi$ to predict structural tokens from the text condition $c$. Unlike left-to-right autoregressive generation, $G_\phi$ uses bidirectional self-attention and iterative masked prediction, so each token position attends to all other positions simultaneously during training, enabling a globally consistent contextual modeling of the sequence.

\noindent\textbf{Training.}
During training, a binary mask $m \in \{0,1\}^N$ is sampled with masking ratio $\rho \in (0, 1]$, where each entry is independently drawn from a Bernoulli distribution with parameter $\rho$. The masked sequence is constructed as:

\begin{equation}
    \tilde{y} = m \odot y + (1-m) \odot y_{\mathrm{mask}},
\end{equation}
with where $y_{\mathrm{mask}}$ denotes a sequence filled the \texttt{[MASK]} token and $\odot$ denotes element-wise multiplication. Taking $\tilde{y}$ and $c$ as inputs, $G_\phi$ predicts the distribution at each position:
\begin{equation}
    p_\phi(y_i \mid \tilde{y}, c).
\end{equation}
The training objectiveis the cross-entropy loss computed only over the masked positions:
\begin{equation}
    \mathcal{L}_{\mathrm{tok}} = -\sum_{i=1}^{N} (1-m_i)\log p_\phi(y_i \mid \tilde{y}, c).
    \label{eq:tok}
\end{equation}

\noindent\textbf{Inference.}
As illustrated in Fig.~\ref{fig:mask}, at $t=0$ all $N$ positions are initialized as \texttt{[MASK]}. At each step $t \in \{1,\ldots,S\}$, $G_\phi$ predicts all masked positions in parallel conditioned on $c$ and the currently visible tokens. The highest-confidence tokens are filled, while the remaining low-confidence positions are re-masked for the next step. The masking ratio $\rho$ is reduced following a cosine annealing schedule:
\begin{equation}
    \rho(t) = \cos\!\left(\frac{t}{S} \cdot \frac{\pi}{2}\right)^2,
\end{equation}
so that $\rho$ decreases from 1 at $t=0$ to 0 at step $S$, gradually revealing the complete index sequence. After $S$ steps, all positions are filled to produce the discrete index sequence $\hat{y}$, which is mapped to the quantized structural prototype $\hat{z}_q = \mathcal{C}[\hat{y}]$. Compared to left-to-right autoregressive generation~\cite{guo2022tm2t}, this iterative masked prediction allows all positions to be decoded in parallel while maintaining global context through bidirectional attention, which is particularly beneficial for generating globally consistent structural token sequences, reducing error accumulation across positions.

\subsection{Conditional Residual Diffusion}
\label{sec:resdiff}

After the discrete structural tokens are predicted, DC-Motion generates the continuous residual component that complements the quantized prototype. Instead of applying diffusion to the full latent representation $z$, we model the scaled residual $r=s(z-z_q)$ produced by DC-VAE. Since $r$ is defined relative to the discrete prototype $z_q$, it represents the component of the continuous latent that is not expressed by the selected codebook entry.

\noindent\textbf{Training.}
During training, DC-VAE provides the ground-truth quantized prototype $z_q$ and residual $r_0=r$. Gaussian noise is gradually added to $r_0$ following a forward diffusion process:
\begin{equation}
    q(r_t \mid r_0) = \mathcal{N}\left(r_t; \sqrt{\bar{\alpha}_t}r_0, (1-\bar{\alpha}_t)I\right),
\end{equation}
or equivalently,
\begin{equation}
    r_t = \sqrt{\bar{\alpha}_t}r_0 + \sqrt{1-\bar{\alpha}_t}\epsilon,
    \quad \epsilon \sim \mathcal{N}(0,I),
\end{equation}
where $\bar{\alpha}_t$ denotes the cumulative noise schedule. A denoising network $\epsilon_\psi$ is trained to predict the injected noise from the noisy residual $r_t$, conditioned on the timestep $t$, the text condition $c$, and the quantized structural representation $z_q$:
\begin{equation}
    \hat{\epsilon} = \epsilon_\psi(r_t,t,c,z_q).
\end{equation}
The training objective is the standard $\epsilon$-prediction loss:
\begin{equation}
    \mathcal{L}_{\mathrm{diff}}
    =
    \mathbb{E}_{t,\epsilon}
    \left[
    \left\|
    \epsilon-\epsilon_\psi(r_t,t,c,z_q)
    \right\|_2^2
    \right].
\label{eq:diff}
\end{equation}

\noindent\textbf{Inference.}
During inference, the discrete index sequence $\hat{y}$ generated by the structural token generator $G_\phi$ is first mapped to its quantized prototype $\hat{z}_q=\mathcal{C}[\hat{y}]$. Starting from Gaussian noise, the residual diffusion model progressively denoises a residual latent $\hat{r}$ conditioned on both the text condition $c$ and the predicted structure $\hat{z}_q$. The final latent is reconstructed as:
\begin{equation}
    \hat{z}=\hat{z}_q+\hat{r}/s,
\end{equation}
and decoded into the final motion sequence:
\begin{equation}
    \hat{x}=D_\theta(\hat{z}).
\label{eq:decode}
\end{equation}
This residual generation process focuses the diffusion model on continuous details not represented by the discrete prototype, rather than requiring it to model the entire motion latent distribution from scratch.
\subsection{Overall Inference}
\label{sec:train_infer}

\begin{algorithm}[t]
\caption{DC-Motion Inference}
\label{alg:inference}
\begin{algorithmic}[1]
\REQUIRE Text condition $c$, codebook $\mathcal{C}$, decoder $D_\theta$, structural token generator $G_\phi$, residual diffusion model $\epsilon_\psi$, masked prediction steps $S$, diffusion steps $T$, residual scaling factor $s$
\ENSURE Generated motion sequence $\hat{x}$
\STATE \textit{// Stage 1: Text-conditioned structural token generation (Sec.~\ref{sec:maskgen})}
\STATE Initialize fully masked sequence $\tilde{y} \leftarrow [\texttt{MASK}]^N$
\FOR{$\tau = 1$ to $S$}
    \STATE Predict $p_\phi(y_i \mid \tilde{y}, c)$ for all masked positions
    \STATE Fill high-confidence positions with predicted tokens
    \STATE Update masking ratio $\rho(\tau)$ using the cosine schedule
\ENDFOR
\STATE Obtain discrete structural index sequence $\hat{y}$
\STATE Retrieve quantized structural prototype $\hat{z}_q = \mathcal{C}[\hat{y}]$
\STATE \textit{// Stage 2: Conditional residual diffusion (Sec.~\ref{sec:resdiff})}
\STATE Sample $\hat{r}_T \sim \mathcal{N}(0, I)$
\FOR{$t = T$ to $1$}
    \STATE $\hat{r}_{t-1} \leftarrow \mathrm{Denoise}(\hat{r}_t, t, c, \hat{z}_q; \epsilon_\psi)$ \hfill $\triangleright$ Eq.~\ref{eq:diff}
\ENDFOR
\STATE \textit{// Decode final motion}
\STATE $\hat{z} = \hat{z}_q + \hat{r}_0/s$
\STATE $\hat{x} = D_\theta(\hat{z})$ \hfill $\triangleright$ Eq.~\ref{eq:decode}
\RETURN $\hat{x}$
\end{algorithmic}
\end{algorithm}

Given a text description $c$, DC-Motion first predicts the discrete structural index sequence $\hat{y}$ using the text-conditioned structural token generator $G_\phi$. The predicted indices are then mapped to the codebook to obtain the quantized structural prototype $\hat{z}_q=\mathcal{C}[\hat{y}]$. Conditioned on both $c$ and $\hat{z}_q$, the residual diffusion model progressively denoises and generates the continuous residual $\hat{r}_0$. Finally, the predicted structure and residual are combined as $\hat{z}=\hat{z}_q+\hat{r}_0/s$ and decoded by $D_\theta$ to obtain the final motion sequence $\hat{x}$. The complete procedure is summarized in Algorithm~\ref{alg:inference}.

\section{Experiments}
\begin{figure*}[t]
    \centering
    \includegraphics[width=0.82\textwidth]{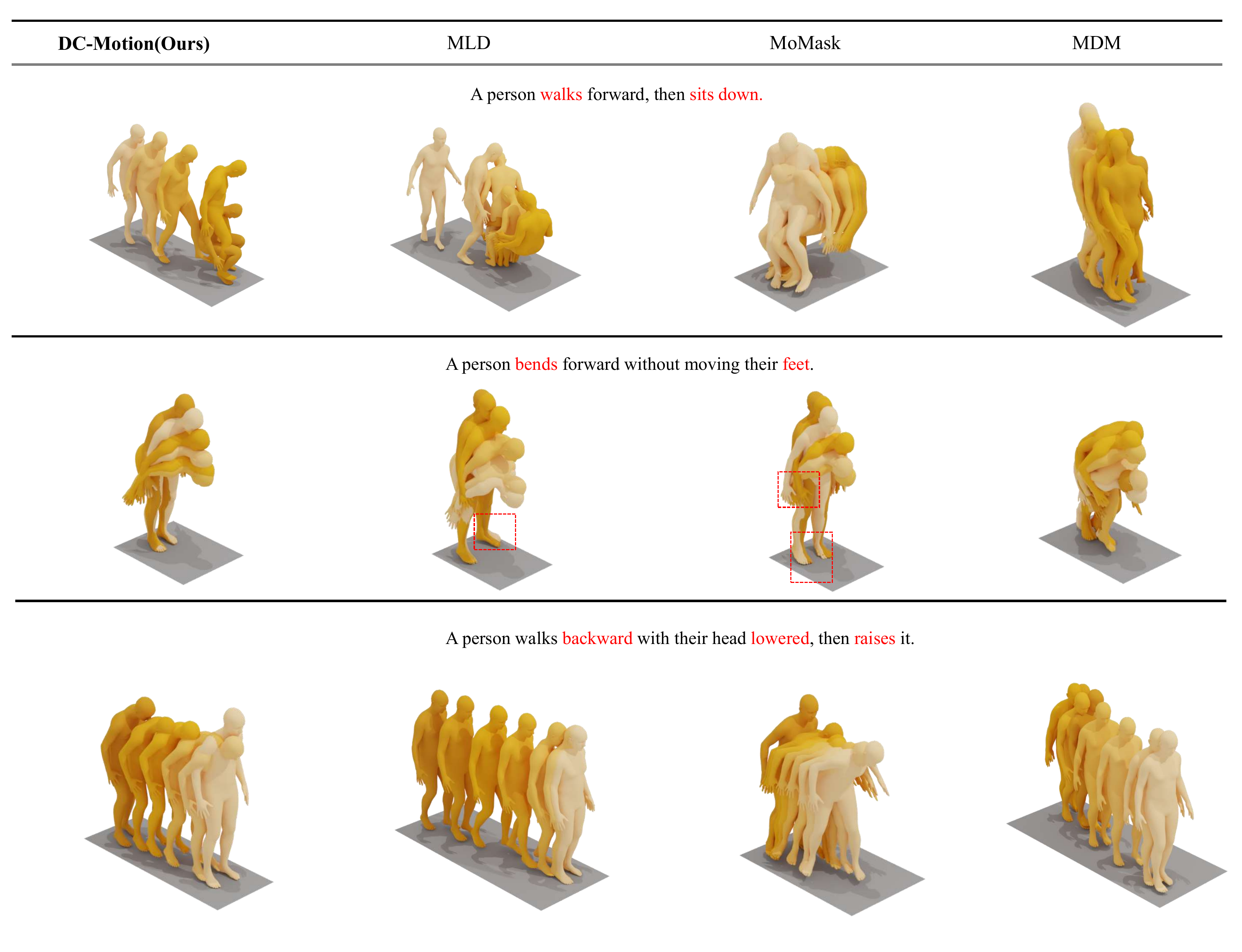}
    \caption{Qualitative comparisons on the HumanML3D test set. Compared with state-of-the-art baselines, DC-Motion better captures fine-grained textual semantics, producing more realistic, coherent, and semantically faithful motions across diverse action categories. Representative failure cases of baseline methods are highlighted by the dashed red boxes.}
    \label{fig:exp}
\end{figure*}
We conduct experiments to evaluate the proposed discrete-continuous factorized framework for text-to-motion generation across motion quality, text-motion alignment, and generation diversity. We introduce the datasets and evaluation metrics (Sec.~\ref{sec:dataset}), implementation details (Sec.~\ref{sec:impl}), followed by quantitative comparisons (Sec.~\ref{sec:results}) and ablation studies (Sec.~\ref{sec:ablation}). We compare DC-Motion against three representative categories of baselines: continuous latent diffusion methods including MDM~\cite{tevet2022human}, MLD~\cite{chen2023executing}, MotionDiffuse~\cite{zhang2024motiondiffuse}, and ReMoDiffuse~\cite{zhang2023remodiffuse}; discrete token generation methods including T2M-GPT~\cite{zhang2023generating} and MotionGPT~\cite{jiang2023motiongpt}; and masked token and residual quantization methods including MMM~\cite{pinyoanuntapong2024mmm} and MoMask~\cite{guo2024momask}. Additional qualitative results are provided in the supplementary material.Further experimental results and analyses are provided in the supplementary material, covering challenging subsets, physical plausibility, intervention analysis, hyperparameter sensitivity, fairness-controlled comparisons, zero-shot generalization, and user study.

\subsection{Datasets and Evaluation Metrics}
\label{sec:dataset}

\noindent\textbf{Datasets.}
We evaluate DC-Motion on two widely used text-to-motion benchmarks: HumanML3D~\cite{guo2022generating} and KIT Motion-Language~\cite{plappert2016kit}. HumanML3D contains 14,616 motion sequences derived from AMASS~\cite{mahmood2019amass}, paired with 44,970 textual descriptions. KIT Motion-Language contains 3,911 motion sequences and 6,353 textual descriptions. Following standard protocols~\cite{tevet2022human, chen2023executing, zhang2024motiondiffuse}, we use the common redundant motion representation that includes joint positions, velocities, rotations, and foot contact labels.

\noindent\textbf{Evaluation Metrics.}
We evaluate generated motions from three aspects: motion quality, text-motion alignment, and generation diversity. For motion quality, we report Fréchet Inception Distance (FID), which measures the distribution distance between generated and real motions in the feature space of a pretrained evaluator; lower values indicate better motion realism. For text-motion alignment, we report R-Precision at Top-1, Top-2, and Top-3, which measures whether the matched text-motion pair can be retrieved among the top-ranked candidates. We also report Multi-Modal Distance (MM Dist), which measures the feature-space distance between each generated motion and its corresponding text condition; lower values indicate better semantic alignment. For diversity, we report Diversity and Multi-Modality. Diversity measures the variation among generated motions across different text descriptions, while Multi-Modality measures the variation among multiple motions generated from the same text prompt.

\subsection{Implementation Details}
\label{sec:impl}

\noindent\textbf{Motion Representation.}
Following MLD~\cite{chen2023executing}, each motion sequence is represented using joint rotations, global positions, velocities, and foot contact signals.

\noindent\textbf{Architecture.}
The encoder $E_\theta$ and decoder $D_\theta$ of DC-VAE share the identical transformer backbone with the MLD tokenizer. Both are implemented as 9-layer transformers with 4 attention heads per layer and standard skip connections. The latent representation is projected into a compact space $z \in \mathbb{R}^{N \times D}$, where the downsampled sequence length is fixed to $N=16$ and the channel dimension to $D=256$. The discrete codebook size is set to $K=2048$.

\noindent\textbf{Optimization and Training Strategy.}
All modules are optimized using AdamW with a fixed learning rate of $1\times10^{-4}$ and no weight decay. Training proceeds in three sequential stages: (1) \textit{DC-VAE Tokenizer Stage}: the autoencoding tokenizer is trained for 6,000 epochs with a batch size of 128; (2) \textit{Structural Token Generator Stage}: the text-conditioned structural token generator is trained for approximately 3,000 epochs with a batch size of 128, while DC-VAE is frozen; (3) \textit{Residual Diffusion Stage}: the residual diffusion model is trained for 3,000 epochs with a batch size of 64, while both DC-VAE and the structural token generator remain frozen throughout.

\noindent\textbf{Diffusion Settings.}
We adopt the standard DDPM formulation for residual diffusion training. The number of diffusion timesteps is set to $T_{\mathrm{train}}=1000$ during training and accelerated to $T_{\mathrm{infer}}=50$ during inference. The noise variance schedule $\{\beta_t\}$ grows linearly from $8.5\times10^{-4}$ to $0.012$, following the configuration in MLD~\cite{chen2023executing}.

\noindent\textbf{Hardware Environment.} The entire three-stage training process is executed on a server equipped with eight NVIDIA RTX 4090 GPUs. Benefiting from our highly efficient decoupled architecture, the evaluation and inference pipelines operate seamlessly on a single consumer-grade GPU.

\subsection{Main Results}
\label{sec:results}
\subsubsection{Quantitative Results.}
We evaluate DC-Motion against three categories of baselines on HumanML3D and KIT Motion-Language. Continuous diffusion baselines include MDM~\cite{tevet2022human}, MLD~\cite{chen2023executing}, MotionDiffuse~\cite{zhang2024motiondiffuse}, and ReMoDiffuse~\cite{zhang2023remodiffuse}. Discrete token generation baselines include TM2T~\cite{guo2022tm2t}, T2M~\cite{guo2022generating}, T2M-GPT~\cite{zhang2023generating}, and MotionGPT~\cite{jiang2023motiongpt}. Masked token and residual quantization baselines include MMM~\cite{pinyoanuntapong2024mmm} and MoMask~\cite{guo2024momask}. All metrics are reported with 95\% confidence intervals over 20 independent runs.

Results on HumanML3D are shown in Tab.~\ref{tab:humanml3d}. DC-Motion achieves the best R-Precision at Top-1, Top-2, and Top-3, and the best MM Dist, outperforming MoMask across all text-motion alignment metrics. This indicates that the discrete structural tokens provide an effective intermediate representation for text-conditioned generation. DC-Motion also achieves the lowest FID among all compared methods, showing that the continuous residual branch reduces distributional discrepancy relative to real motions. In contrast to MoMask, which approximates residual details with additional discrete codebook levels, DC-Motion retains the residual as a continuous latent and models it with a conditional diffusion process. The improvement in both FID and R-Precision over MoMask suggests that this discrete-continuous factorization benefits both motion quality and text alignment simultaneously.

Results on KIT Motion-Language are shown in Tab.~\ref{tab:kit}. DC-Motion achieves the best R-Precision and MM Dist, and the lowest FID among all methods, which is consistent with the trends observed on the HumanML3D results. The performance gap between DC-Motion and ReMoDiffuse on FID is smaller on KIT than on HumanML3D, which may be attributed to the smaller scale of the KIT and its relatively limited diversity.
\begin{figure*}[!t]
\centering
\includegraphics[width=0.82\textwidth]{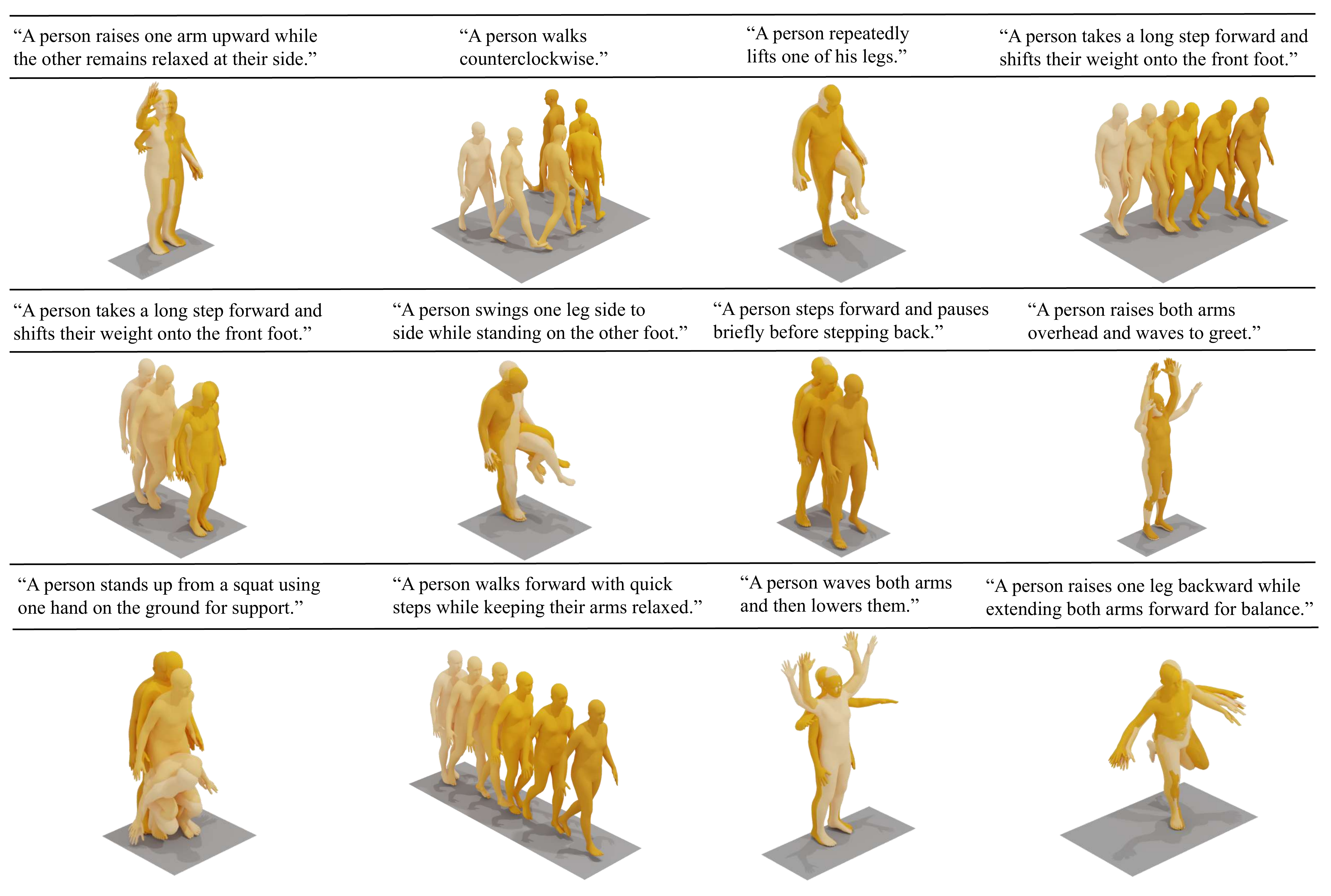}
\caption{Qualitative results of complex multi-stage motion generation. Each row shows a text prompt with key frames sampled at uniform intervals.}
\label{fig:append}
\end{figure*}

\subsubsection{Qualitative Results.}
Fig.~\ref{fig:exp} compares DC-Motion with MDM, MLD, and MoMask on representative text prompts. MDM often fails to complete sequential actions, MLD may produce artifacts such as foot sliding, and MoMask can suffer from stiff transitions caused by discretization boundaries. In contrast, DC-Motion generates smoother and more semantically accurate motions, consistent with its quantitative gains in FID and R-Precision. Fig.~\ref{fig:append} further shows complex multi-stage examples with uniformly sampled key frames. DC-Motion preserves the global action layout across consecutive stages such as walking, turning, jumping, and landing, while maintaining smooth local dynamics and plausible variations in body orientation, limb swing, and transition timing. These results support the proposed decomposition, where discrete structural tokens model coarse temporal organization and phase transitions, and the continuous residual branch refines fine-grained joint trajectories beyond the finite codebook representation.

\begin{table*}[t]
\centering
\scriptsize
\caption{Quantitative comparison on the HumanML3D dataset. \textbf{Bold} and \underline{underline} indicate the best and second-best results. DC-Motion achieves the best R-Precision and FID among all compared methods.}
\setlength{\tabcolsep}{4pt}
\begin{tabular*}{\textwidth}{@{\extracolsep{\fill}}lcccccc}
\toprule
\multirow{2}{*}{Methods} & \multicolumn{3}{c}{R Precision $\uparrow$} & \multirow{2}{*}{FID $\downarrow$} & \multirow{2}{*}{MM Dist $\downarrow$} & \multirow{2}{*}{MModality $\uparrow$} \\
\cmidrule{2-4}
 & Top 1 & Top 2 & Top 3 & & & \\
\midrule
Real & $0.511^{\pm.003}$ & $0.703^{\pm.003}$ & $0.797^{\pm.002}$ & $0.002^{\pm.000}$ & $2.974^{\pm.008}$ & - \\
\midrule
TM2T~\cite{guo2022tm2t} & $0.424^{\pm.003}$ & $0.618^{\pm.003}$ & $0.729^{\pm.002}$ & $1.501^{\pm.017}$ & $3.467^{\pm.011}$ & $2.424^{\pm.093}$ \\
T2M~\cite{guo2022generating} & $0.455^{\pm.003}$ & $0.636^{\pm.003}$ & $0.736^{\pm.002}$ & $1.087^{\pm.021}$ & $3.347^{\pm.008}$ & $2.219^{\pm.074}$ \\
MDM~\cite{tevet2022human} & - & - & $0.611^{\pm.007}$ & $0.544^{\pm.044}$ & $5.566^{\pm.027}$ & $\mathbf{2.799^{\pm.072}}$ \\
MLD~\cite{chen2023executing} & $0.481^{\pm.003}$ & $0.673^{\pm.003}$ & $0.772^{\pm.002}$ & $0.473^{\pm.013}$ & $3.196^{\pm.010}$ & $2.413^{\pm.079}$ \\
MotionDiffuse~\cite{zhang2024motiondiffuse} & $0.491^{\pm.001}$ & $0.681^{\pm.001}$ & $0.782^{\pm.001}$ & $0.630^{\pm.001}$ & $3.113^{\pm.001}$ & $1.553^{\pm.042}$ \\
T2M-GPT~\cite{zhang2023generating} & $0.492^{\pm.003}$ & $0.679^{\pm.002}$ & $0.775^{\pm.002}$ & $0.141^{\pm.005}$ & $3.121^{\pm.009}$ & $1.831^{\pm.048}$ \\
MotionGPT~\cite{jiang2023motiongpt} & $0.492^{\pm.003}$ & $0.681^{\pm.003}$ & $0.778^{\pm.002}$ & $0.232^{\pm.008}$ & $3.096^{\pm.008}$ & $2.008^{\pm.084}$ \\
ReMoDiffuse~\cite{zhang2023remodiffuse} & $0.510^{\pm.005}$ & $0.698^{\pm.006}$ & $0.795^{\pm.004}$ & $0.103^{\pm.004}$ & $2.974^{\pm.016}$ & $1.795^{\pm.043}$ \\
MMM~\cite{pinyoanuntapong2024mmm} & $0.504^{\pm.003}$ & $0.696^{\pm.003}$ & $0.794^{\pm.004}$ & $0.080^{\pm.003}$ & $3.091^{\pm.007}$ & $1.164^{\pm.041}$ \\
MoMask~\cite{guo2024momask} & \underline{$0.521^{\pm.002}$} & \underline{$0.713^{\pm.002}$} & \underline{$0.807^{\pm.002}$} & \underline{$0.045^{\pm.002}$} & \underline{$2.958^{\pm.008}$} & $1.241^{\pm.040}$ \\
\textbf{DC-Motion (Ours)} & $\mathbf{0.528^{\pm.002}}$ & $\mathbf{0.720^{\pm.002}}$ & $\mathbf{0.812^{\pm.002}}$ & $\mathbf{0.041^{\pm.002}}$ & $\mathbf{2.945^{\pm.008}}$ & \underline{$2.450^{\pm.070}$} \\
\bottomrule
\end{tabular*}
\label{tab:humanml3d}
\end{table*}

\begin{table*}[t]
\centering
\scriptsize
\caption{Quantitative comparison on the KIT Motion-Language dataset. \textbf{Bold} and \underline{underline} indicate the best and second-best results. DC-Motion achieves the best R-Precision, MM Dist, and FID among all compared methods.}
\setlength{\tabcolsep}{4pt}
\begin{tabular*}{\textwidth}{@{\extracolsep{\fill}}lcccccc}
\toprule
\multirow{2}{*}{Methods} & \multicolumn{3}{c}{R Precision $\uparrow$} & \multirow{2}{*}{FID $\downarrow$} & \multirow{2}{*}{MM Dist $\downarrow$} & \multirow{2}{*}{MModality $\uparrow$} \\
\cmidrule{2-4}
 & Top 1 & Top 2 & Top 3 & & & \\
\midrule
TM2T~\cite{guo2022tm2t} & $0.280^{\pm.005}$ & $0.463^{\pm.006}$ & $0.587^{\pm.005}$ & $3.599^{\pm.153}$ & $4.591^{\pm.026}$ & $\mathbf{3.292^{\pm.081}}$ \\
T2M~\cite{guo2022generating} & $0.361^{\pm.005}$ & $0.559^{\pm.007}$ & $0.681^{\pm.007}$ & $3.022^{\pm.107}$ & $3.488^{\pm.028}$ & $2.052^{\pm.107}$ \\
MDM~\cite{tevet2022human} & - & - & $0.396^{\pm.004}$ & $0.497^{\pm.021}$ & $9.191^{\pm.022}$ & $1.907^{\pm.214}$ \\
MLD~\cite{chen2023executing} & $0.390^{\pm.008}$ & $0.609^{\pm.008}$ & $0.734^{\pm.007}$ & $0.404^{\pm.027}$ & $3.204^{\pm.027}$ & \underline{$2.192^{\pm.071}$} \\
MotionDiffuse~\cite{zhang2024motiondiffuse} & $0.417^{\pm.004}$ & $0.621^{\pm.004}$ & $0.739^{\pm.004}$ & $1.954^{\pm.062}$ & $2.958^{\pm.005}$ & $0.730^{\pm.013}$ \\
T2M-GPT~\cite{zhang2023generating} & $0.416^{\pm.006}$ & $0.627^{\pm.006}$ & $0.745^{\pm.006}$ & $0.514^{\pm.029}$ & $3.007^{\pm.023}$ & $1.570^{\pm.039}$ \\
MotionGPT~\cite{jiang2023motiongpt} & $0.366^{\pm.006}$ & $0.558^{\pm.007}$ & $0.680^{\pm.006}$ & $0.510^{\pm.016}$ & $3.527^{\pm.019}$ & $2.328^{\pm.097}$ \\
ReMoDiffuse~\cite{zhang2023remodiffuse} & $0.427^{\pm.014}$ & $0.641^{\pm.004}$ & $0.765^{\pm.055}$ & \underline{$0.155^{\pm.006}$} & $2.814^{\pm.012}$ & $1.239^{\pm.028}$ \\
MMM~\cite{pinyoanuntapong2024mmm} & $0.404^{\pm.006}$ & $0.615^{\pm.006}$ & $0.738^{\pm.005}$ & $0.217^{\pm.009}$ & $3.147^{\pm.021}$ & $1.204^{\pm.031}$ \\
MoMask~\cite{guo2024momask} & \underline{$0.433^{\pm.007}$} & \underline{$0.656^{\pm.005}$} & \underline{$0.781^{\pm.005}$} & $0.204^{\pm.011}$ & \underline{$2.779^{\pm.022}$} & $1.131^{\pm.043}$ \\
\textbf{DC-Motion (Ours)} & $\mathbf{0.442^{\pm.005}}$ & $\mathbf{0.665^{\pm.005}}$ & $\mathbf{0.788^{\pm.005}}$ & $\mathbf{0.148^{\pm.008}}$ & $\mathbf{2.745^{\pm.018}}$ & $1.950^{\pm.060}$ \\
\bottomrule
\end{tabular*}
\label{tab:kit}
\end{table*}

\subsection{Ablation Study}
\label{sec:ablation}

To examine the role of each component in DC-Motion, we conduct ablation studies on the HumanML3D dataset. Unless otherwise specified, all variants use the same training protocol, evaluator, and inference settings. The ablations are designed to answer four questions: (1) whether the discrete-continuous tokenizer provides a better representation than purely discrete or purely continuous tokenizers; (2) whether diffusion should be applied to the full latent space or only to the quantization residual; (3) which conditions are necessary for residual diffusion; and (4) whether iterative masked prediction is preferable to left-to-right autoregressive generation for structural token prediction in the compact discrete token space.

\subsubsection{Effect of Discrete-Continuous Tokenization.}
DC-VAE is the representation foundation of DC-Motion. To evaluate its role, we compare it with standard tokenizer baselines, including VQ-VAE, RVQ, and a continuous VAE, as well as internal variants of DC-VAE. The results are shown in Tab.~\ref{tab:dcvae_ablation}.

The baseline tokenizers reveal a reconstruction-generation trade-off. The continuous VAE obtains low reconstruction error but its unstructured latent space is less suited for text-conditioned structural generation, resulting in higher generation FID. VQ-VAE provides a compact discrete space that benefits generation but suffers from larger reconstruction error because fine-grained details are forced into a finite codebook. RVQ reduces this reconstruction gap by stacking additional discrete residual levels, but all residual information remains in discrete form.

DC-VAE alleviates this trade-off by retaining the first-level quantization residual as a continuous latent. Removing the continuous residual branch (w/o Residual) substantially increases reconstruction MPJPE, confirming that the residual branch preserves motion details not captured by the codebook. Removing the discrete branch (w/o Discrete) degrades generation FID, showing that the discrete prototype provides a useful structural abstraction for text-conditioned generation. These results support the core design of DC-Motion: the discrete and continuous components are complementary.

\begin{table}[t]
\centering
\scriptsize
\caption{Ablation study of DC-VAE on HumanML3D. Bold and underline indicate the best and second-best results in each column.}
\label{tab:dcvae_ablation}
\setlength{\tabcolsep}{2.5pt}
\renewcommand{\arraystretch}{1.05}
\begin{tabular*}{\columnwidth}{@{\extracolsep{\fill}}lcccc@{}}
\toprule
\multirow{2}{*}{Method} &
\multicolumn{2}{c}{Recon.} &
\multicolumn{2}{c}{Gen.} \\
\cmidrule(lr){2-3}\cmidrule(lr){4-5}
& FID$\downarrow$ & MPJPE$\downarrow$ & FID$\downarrow$ & MM Dist$\downarrow$ \\
\midrule
VQ-VAE & $0.141$ & $58.0$ & $0.151$ & $3.121$ \\
RVQ & $0.051$ & $29.5$ & $0.059$ & $2.957$ \\
Cont. VAE & $0.083$ & $\mathbf{22.3}$ & $0.083$ & $3.102$ \\
\midrule
w/o Discrete & $\mathbf{0.012}$ & \underline{$23.1$} & $0.081$ & $3.089$ \\
w/o Residual & $0.041$ & $41.7$ & $0.072$ & $2.991$ \\
w/o Skip & $0.021$ & $31.2$ & $0.059$ & $2.973$ \\
$D$=128 & $0.018$ & $28.4$ & $0.054$ & $2.962$ \\
$D$=512 & \underline{$0.013$} & $24.9$ & $0.049$ & $2.951$ \\
$K$=1024 & $0.019$ & $30.1$ & $0.053$ & $2.968$ \\
$K$=4096 & \underline{$0.013$} & $25.2$ & \underline{$0.047$} & \underline{$2.949$} \\
\midrule
DC-VAE & $0.014$ & $25.8$ & $\mathbf{0.041}$ & $\mathbf{2.945}$ \\
\bottomrule
\end{tabular*}
\end{table}

\subsubsection{Effect of Residual Diffusion Target.}
We first study the choice of diffusion target. Since DC-VAE decomposes the motion latent into a discrete prototype $z_q$ and a continuous residual $r=s(z-z_q)$, the diffusion model can be trained on different target spaces: raw motion $x$, the full latent $z$, or the residual $r$. Raw motion diffusion directly operates on frame-level motion sequences and therefore has higher computational cost. Full-latent diffusion models the entire continuous latent, which requires the denoiser to jointly learn structural layout and local dynamics. In contrast, residual diffusion models only the continuous complement of the discrete prototype, conditioned on both the text $c$ and the quantized structure $z_q$.

As shown in Tab.~\ref{tab:diff_target}, residual diffusion achieves the best generation quality among the compared target spaces. Compared with full-latent diffusion, residual diffusion improves both FID and MM Dist, indicating that the residual target is easier to model than the full latent distribution. Compared with raw motion diffusion, residual diffusion operates in a compact latent-token space and avoids the computational cost of frame-level denoising. These results support the design choice of applying diffusion to the quantization residual rather than to the entire motion representation.

\begin{table}[t]
\centering
\scriptsize
\caption{Ablation study of diffusion target space on HumanML3D.}
\label{tab:diff_target}
\setlength{\tabcolsep}{3pt}
\renewcommand{\arraystretch}{1.05}
\begin{tabular*}{\columnwidth}{@{\extracolsep{\fill}}lcccc@{}}
\toprule
Target & FID$\downarrow$ & MM Dist$\downarrow$ & Time$\downarrow$ & VRAM$\downarrow$ \\
\midrule
Raw motion $x$ & $0.121$ & $3.25$ & $2.87\times$ & $3.12\times$ \\
Full latent $z$ & $0.081$ & $3.089$ & $1.00\times$ & $1.00\times$ \\
Residual $r$ & $\mathbf{0.041}$ & $\mathbf{2.945}$ & $1.00\times$ & $1.00\times$ \\
\bottomrule
\end{tabular*}
\end{table}

\subsubsection{Effect of Residual Diffusion Conditioning.}
We further analyze which conditions are necessary for residual diffusion. The residual $r=s(z-z_q)$ is defined relative to the quantized prototype $z_q$, while the text condition $c$ provides semantic guidance for local motion details. To examine their individual roles and mutual contributions, we compare three variants: conditioning the residual diffusion model on text only, on structure only, and on both text and structure simultaneously.

As shown in Tab.~\ref{tab:res_cond}, removing the structural condition $z_q$ weakens the denoiser because the residual is no longer anchored to the discrete prototype from which it is defined. In this case, the model must infer both the residual reference point and the local dynamics from text alone. Removing the text condition also degrades text-motion alignment, suggesting that local residual details still benefit from semantic guidance. The full model, conditioned on both $c$ and $z_q$, achieves the best overall performance. This verifies that residual diffusion is most effective when it is tied to the discrete-continuous factorization produced by DC-VAE.

\begin{table}[t]
\centering
\scriptsize
\caption{Ablation study of residual diffusion conditioning on HumanML3D}
\label{tab:res_cond}
\setlength{\tabcolsep}{3pt}
\renewcommand{\arraystretch}{1.05}
\begin{tabular*}{\columnwidth}{@{\extracolsep{\fill}}lccc@{}}
\toprule
Condition & FID$\downarrow$ & MM Dist$\downarrow$ & R@1$\uparrow$ \\
\midrule
Text only & $0.062$ & $3.06$ & $0.492$ \\
Structure only & $0.058$ & $3.03$ & $0.3089$ \\
Text + Structure & $\mathbf{0.041}$ & $\mathbf{2.945}$ & $\mathbf{0.528}$ \\
\bottomrule
\end{tabular*}
\end{table}

\subsubsection{Effect of Structural Token Generation.}
We compare the text-conditioned structural token generator with a standard left-to-right autoregressive generator. The results are summarized in Tab.~\ref{tab:gen_ablation}. The autoregressive generator predicts tokens sequentially and can accumulate prediction errors along the sequence, leading to inconsistent global structure. The structural token generator instead refines low-confidence positions iteratively using bidirectional context, which is better suited for generating a compact and globally coherent structural token sequence that must be consistent with the text condition.

The structural token generator achieves better R-Precision and FID than the autoregressive baseline, with larger gains observed on the long-sequence and complex-prompt subsets. This indicates that structural tokens benefit from global context during prediction, consistent with the role of discrete tokens as a compact action layout rather than frame-level detail.

\begin{table}[t]
\centering
\scriptsize
\caption{Ablation study of structural token generation on HumanML3D.}
\label{tab:gen_ablation}
\setlength{\tabcolsep}{3pt}
\renewcommand{\arraystretch}{1.05}
\begin{tabular*}{\columnwidth}{@{\extracolsep{\fill}}lcccc@{}}
\toprule
\multirow{2}{*}{Generator} &
\multirow{2}{*}{FID$\downarrow$} &
\multicolumn{3}{c}{R@1$\uparrow$} \\
\cmidrule(lr){3-5}
& & Overall & Long & Complex \\
\midrule
Autoregressive & $0.213$ & $0.611$ & $0.574$ & $0.552$ \\
Masked & $\mathbf{0.182}$ & $\mathbf{0.642}$ & $\mathbf{0.623}$ & $\mathbf{0.601}$ \\
\bottomrule
\end{tabular*}
\end{table}

\section{Conclusion}
In this paper, we proposed DC-Motion, a discrete-continuous factorized framework for text-to-motion generation. DC-Motion represents motion with discrete structural tokens for action layout and continuous residual latents for details not captured by the codebook. A text-conditioned structural token generator predicts the discrete structure, while a residual diffusion model generates the continuous complement conditioned on both the text and structure. Experiments on HumanML3D and KIT Motion-Language show that this factorization achieves a practical balance between motion quality and text-motion alignment, and the ablation studies verify the contribution of each component.

\section*{Acknowledgments}
This work was supported by the NSFC Regional Innovation and Development Joint Fund under Grant U25A20537, and the National Key Research and Development Program of China under Grant 2024YFC3015600. The numerical calculation is supported by the super-computing system in the Super-computing center of Wuhan University.


\bibliographystyle{IEEEtran}
\bibliography{references}

@String{Computer = "{IEEE} Computer" }

@String{Springer = "Springer-Verlag" }

@inproceedings{ahuja2019language2pose,
  title={Language2pose: Natural language grounded pose forecasting},
  author={Ahuja, Chaitanya and Morency, Louis-Philippe},
  booktitle={2019 International conference on 3D vision (3DV)},
  pages={719--728},
  year={2019},
  organization={IEEE}
}

@inproceedings{kim2023flame,
  title={Flame: Free-form language-based motion synthesis \& editing},
  author={Kim, Jihoon and Kim, Jiseob and Choi, Sungjoon},
  booktitle={Proceedings of the AAAI conference on artificial intelligence},
  volume={37},
  number={7},
  pages={8255--8263},
  year={2023}
}

@inproceedings{petrovich2022temos,
  title={Temos: Generating diverse human motions from textual descriptions},
  author={Petrovich, Mathis and Black, Michael J and Varol, G{\"u}l},
  booktitle={European conference on computer vision},
  pages={480--497},
  year={2022},
  organization={Springer}
}

@article{lu2023humantomato,
  title={Humantomato: Text-aligned whole-body motion generation},
  author={Lu, Shunlin and Chen, Ling-Hao and Zeng, Ailing and Lin, Jing and Zhang, Ruimao and Zhang, Lei and Shum, Heung-Yeung},
  journal={arXiv preprint arXiv:2310.12978},
  year={2023}
}

@article{zhang2024motiondiffuse,
  title={Motiondiffuse: Text-driven human motion generation with diffusion model},
  author={Zhang, Mingyuan and Cai, Zhongang and Pan, Liang and Hong, Fangzhou and Guo, Xinying and Yang, Lei and Liu, Ziwei},
  journal={IEEE transactions on pattern analysis and machine intelligence},
  volume={46},
  number={6},
  pages={4115--4128},
  year={2024},
  publisher={IEEE}
}

@article{tevet2022human,
  title={Human motion diffusion model},
  author={Tevet, Guy and Raab, Sigal and Gordon, Brian and Shafir, Yonatan and Cohen-Or, Daniel and Bermano, Amit H},
  journal={arXiv preprint arXiv:2209.14916},
  year={2022}
}

@inproceedings{chen2023executing,
  title={Executing your commands via motion diffusion in latent space},
  author={Chen, Xin and Jiang, Biao and Liu, Wen and Huang, Zilong and Fu, Bin and Chen, Tao and Yu, Gang},
  booktitle={Proceedings of the IEEE/CVF conference on computer vision and pattern recognition},
  pages={18000--18010},
  year={2023}
}

@article{yu2026causal,
  title={Causal Motion Diffusion Models for Autoregressive Motion Generation},
  author={Yu, Qing and Watanabe, Akihisa and Fujiwara, Kent},
  journal={arXiv preprint arXiv:2602.22594},
  year={2026}
}

@article{li2026llamo,
  title={LLaMo: Scaling Pretrained Language Models for Unified Motion Understanding and Generation with Continuous Autoregressive Tokens},
  author={Li, Zekun and An, Sizhe and Tang, Chengcheng and Guo, Chuan and Shugurov, Ivan and Zhang, Linguang and Zhao, Amy and Sridhar, Srinath and Tao, Lingling and Mittal, Abhay},
  journal={arXiv preprint arXiv:2602.12370},
  year={2026}
}

@inproceedings{guo2022tm2t,
  title={Tm2t: Stochastic and tokenized modeling for the reciprocal generation of 3d human motions and texts},
  author={Guo, Chuan and Zuo, Xinxin and Wang, Sen and Cheng, Li},
  booktitle={European Conference on Computer Vision},
  pages={580--597},
  year={2022},
  organization={Springer}
}

@article{wang2026temporal,
  title={Temporal consistency-aware text-to-motion generation},
  author={Wang, Hongsong and Yan, Wenjing and Lai, Qiuxia and Geng, Xin},
  journal={Visual Intelligence},
  volume={4},
  number={1},
  pages={7},
  year={2026},
  publisher={Springer}
}

@inproceedings{petrovich2021action,
  title={Action-conditioned 3d human motion synthesis with transformer vae},
  author={Petrovich, Mathis and Black, Michael J and Varol, G{\"u}l},
  booktitle={Proceedings of the IEEE/CVF international conference on computer vision},
  pages={10985--10995},
  year={2021}
}

@inproceedings{ahn2018text2action,
  title={Text2action: Generative adversarial synthesis from language to action},
  author={Ahn, Hyemin and Ha, Timothy and Choi, Yunho and Yoo, Hwiyeon and Oh, Songhwai},
  booktitle={2018 IEEE International Conference on Robotics and Automation (ICRA)},
  pages={5915--5920},
  year={2018},
  organization={IEEE}
}

@inproceedings{zhang2023generating,
  title={Generating human motion from textual descriptions with discrete representations},
  author={Zhang, Jianrong and Zhang, Yangsong and Cun, Xiaodong and Zhang, Yong and Zhao, Hongwei and Lu, Hongtao and Shen, Xi and Shan, Ying},
  booktitle={Proceedings of the IEEE/CVF conference on computer vision and pattern recognition},
  pages={14730--14740},
  year={2023}
}

@article{jiang2023motiongpt,
  title={Motiongpt: Human motion as a foreign language},
  author={Jiang, Biao and Chen, Xin and Liu, Wen and Yu, Jingyi and Yu, Gang and Chen, Tao},
  journal={Advances in Neural Information Processing Systems},
  volume={36},
  pages={20067--20079},
  year={2023}
}

@inproceedings{guo2024momask,
  title={Momask: Generative masked modeling of 3d human motions},
  author={Guo, Chuan and Mu, Yuxuan and Javed, Muhammad Gohar and Wang, Sen and Cheng, Li},
  booktitle={Proceedings of the IEEE/CVF Conference on Computer Vision and Pattern Recognition},
  pages={1900--1910},
  year={2024}
}

@article{van2017neural,
  title={Neural discrete representation learning},
  author={Van Den Oord, Aaron and Vinyals, Oriol and others},
  journal={Advances in neural information processing systems},
  volume={30},
  year={2017}
}

@inproceedings{guo2022generating,
  title={Generating diverse and natural 3d human motions from text},
  author={Guo, Chuan and Zou, Shihao and Zuo, Xinxin and Wang, Sen and Ji, Wei and Li, Xingyu and Cheng, Li},
  booktitle={Proceedings of the IEEE/CVF conference on computer vision and pattern recognition},
  pages={5152--5161},
  year={2022}
}

@article{plappert2016kit,
  title={The kit motion-language dataset},
  author={Plappert, Matthias and Mandery, Christian and Asfour, Tamim},
  journal={Big data},
  volume={4},
  number={4},
  pages={236--252},
  year={2016},
  publisher={SAGE Publications Sage CA: Los Angeles, CA}
}

@inproceedings{mahmood2019amass,
  title={AMASS: Archive of motion capture as surface shapes},
  author={Mahmood, Naureen and Ghorbani, Nima and Troje, Nikolaus F and Pons-Moll, Gerard and Black, Michael J},
  booktitle={Proceedings of the IEEE/CVF international conference on computer vision},
  pages={5442--5451},
  year={2019}
}

@article{zhang2024modular,
  title={A modular neural motion retargeting system decoupling skeleton and shape perception},
  author={Zhang, Jiaxu and Tu, Zhigang and Weng, Junwu and Yuan, Junsong and Du, Bo},
  journal={IEEE Transactions on Pattern Analysis and Machine Intelligence},
  volume={46},
  number={10},
  pages={6889--6904},
  year={2024},
  publisher={IEEE}
}

@inproceedings{chen2025motion,
  title={Motion-example-controlled Co-speech Gesture Generation Leveraging Large Language Models},
  author={Chen, Bohong and Li, Yumeng and Zheng, Youyi and Ding, Yao-Xiang and Zhou, Kun},
  booktitle={Proceedings of the Special Interest Group on Computer Graphics and Interactive Techniques Conference Conference Papers},
  pages={1--12},
  year={2025}
}

@inproceedings{zeng2025light,
  title={Light-t2m: A lightweight and fast model for text-to-motion generation},
  author={Zeng, Ling-An and Huang, Guohong and Wu, Gaojie and Zheng, Wei-Shi},
  booktitle={Proceedings of the AAAI Conference on Artificial Intelligence},
  volume={39},
  number={9},
  pages={9797--9805},
  year={2025}
}

@inproceedings{zhang2023remodiffuse,
  title={Remodiffuse: Retrieval-augmented motion diffusion model},
  author={Zhang, Mingyuan and Guo, Xinying and Pan, Liang and Cai, Zhongang and Hong, Fangzhou and Li, Huirong and Yang, Lei and Liu, Ziwei},
  booktitle={Proceedings of the IEEE/CVF International Conference on Computer Vision},
  pages={364--373},
  year={2023}
}

@inproceedings{zuffi2018lions,
  title={Lions and tigers and bears: Capturing non-rigid, 3d, articulated shape from images},
  author={Zuffi, Silvia and Kanazawa, Angjoo and Black, Michael J},
  booktitle={Proceedings of the IEEE conference on Computer Vision and Pattern Recognition},
  pages={3955--3963},
  year={2018}
}

@inproceedings{dang2026segmo,
  title={SegMo: Segment-aligned text to 3D human motion generation},
  author={Dang, Bowen and Wu, Lin and Yang, Xiaohang and Yuan, Zheng and Chen, Zhixiang},
  booktitle={Proceedings of the IEEE/CVF Winter Conference on Applications of Computer Vision},
  pages={6946--6955},
  year={2026}
}

@inproceedings{gong2026diffusion,
  title={Diffusion implicit policy for unpaired scene-aware motion synthesis},
  author={Gong, Jingyu and Zhang, Chong and Liu, Fengqi and Fan, Ke and Zhou, Qianyu and Tan, Xin and Zhang, Zhizhong and Xie, Yuan},
  booktitle={Proceedings of the AAAI Conference on Artificial Intelligence},
  volume={40},
  number={6},
  pages={4257--4265},
  year={2026}
}

@article{gang2025strong,
  title={Strong and Controllable 3D Motion Generation},
  author={Gang, Canxuan},
  journal={arXiv preprint arXiv:2501.18726},
  year={2025}
}

@inproceedings{athanasiou2022teach,
  title={Teach: Temporal action composition for 3d humans},
  author={Athanasiou, Nikos and Petrovich, Mathis and Black, Michael J and Varol, G{\"u}l},
  booktitle={2022 International Conference on 3D Vision (3DV)},
  pages={414--423},
  year={2022},
  organization={IEEE}
}

@article{taylor2006modeling,
  title={Modeling human motion using binary latent variables},
  author={Taylor, Graham W and Hinton, Geoffrey E and Roweis, Sam},
  journal={Advances in neural information processing systems},
  volume={19},
  year={2006}
}

@inproceedings{zuffi20173d,
  title={3D menagerie: Modeling the 3D shape and pose of animals},
  author={Zuffi, Silvia and Kanazawa, Angjoo and Jacobs, David W and Black, Michael J},
  booktitle={Proceedings of the IEEE conference on computer vision and pattern recognition},
  pages={6365--6373},
  year={2017}
}

@inproceedings{niewiadomski2025generative,
  title={Generative zoo},
  author={Niewiadomski, Tomasz and Yiannakidis, Anastasios and Cuevas-Velasquez, Hanz and Sanyal, Soubhik and Black, Michael J and Zuffi, Silvia and Kulits, Peter},
  booktitle={Proceedings of the IEEE/CVF International Conference on Computer Vision},
  pages={8492--8502},
  year={2025}
}

@inproceedings{pinyoanuntapong2024mmm,
  title={Mmm: Generative masked motion model},
  author={Pinyoanuntapong, Ekkasit and Wang, Pu and Lee, Minwoo and Chen, Chen},
  booktitle={Proceedings of the IEEE/CVF Conference on Computer Vision and Pattern Recognition},
  pages={1546--1555},
  year={2024}
}

@article{dong2026motionflow,
  title={MotionFlow: Efficient Motion Generation with Latent Flow Matching},
  author={Dong, Kun and Xue, Jian and Lan, Xing and Liu, Qingyuan and Lu, Ke},
  journal={IEEE Transactions on Multimedia},
  year={2026},
  publisher={IEEE}
}

@inproceedings{zhang2023skinned,
  title={Skinned motion retargeting with residual perception of motion semantics \& geometry},
  author={Zhang, Jiaxu and Weng, Junwu and Kang, Di and Zhao, Fang and Huang, Shaoli and Zhe, Xuefei and Bao, Linchao and Shan, Ying and Wang, Jue and Tu, Zhigang},
  booktitle={Proceedings of the IEEE/CVF Conference on Computer Vision and Pattern Recognition},
  pages={13864--13872},
  year={2023}
}

@inproceedings{zhang2025semtalk,
  title={Semtalk: Holistic co-speech motion generation with frame-level semantic emphasis},
  author={Zhang, Xiangyue and Li, Jianfang and Zhang, Jiaxu and Dang, Ziqiang and Ren, Jianqiang and Bo, Liefeng and Tu, Zhigang},
  booktitle={Proceedings of the IEEE/CVF International Conference on Computer Vision},
  pages={13761--13771},
  year={2025}
}

@inproceedings{zhang2022mixste,
  title={Mixste: Seq2seq mixed spatio-temporal encoder for 3d human pose estimation in video},
  author={Zhang, Jinlu and Tu, Zhigang and Yang, Jianyu and Chen, Yujin and Yuan, Junsong},
  booktitle={Proceedings of the IEEE/CVF conference on computer vision and pattern recognition},
  pages={13232--13242},
  year={2022}
}

@article{tu2022joint,
  title={Joint-bone fusion graph convolutional network for semi-supervised skeleton action recognition},
  author={Tu, Zhigang and Zhang, Jiaxu and Li, Hongyan and Chen, Yujin and Yuan, Junsong},
  journal={IEEE Transactions on Multimedia},
  volume={25},
  pages={1819--1831},
  year={2022},
  publisher={IEEE}
}

@article{cai2026coordinating,
  title={Coordinating Multiple Conditions for Trajectory-Controlled Human Motion Generation},
  author={Cai, Deli and Ma, Haoyang and Ding, Changxing},
  journal={IEEE Transactions on Multimedia},
  year={2026},
  publisher={IEEE}
}

@article{gao2025jointly,
  title={Jointly understand your command and intention: Reciprocal co-evolution between scene-aware 3d human motion synthesis and analysis},
  author={Gao, Xuehao and Yang, Yang and Du, Shaoyi and Qi, Guo-Jun and Han, Junwei},
  journal={IEEE Transactions on Multimedia},
  year={2025},
  publisher={IEEE}
}

@article{yu2024divdiff,
  title={Divdiff: A conditional diffusion model for diverse human motion prediction},
  author={Yu, Hua and Hou, Yaqing and Pei, Wenbin and Ong, Yew-Soon and Zhang, Qiang},
  journal={IEEE Transactions on Multimedia},
  volume={27},
  pages={1848--1859},
  year={2024},
  publisher={IEEE}
}

@article{wang2024cross,
  title={Cross-modal quantization for co-speech gesture generation},
  author={Wang, Zheng and Zhang, Wei and Ye, Long and Zeng, Dan and Mei, Tao},
  journal={IEEE Transactions on Multimedia},
  volume={26},
  pages={10251--10263},
  year={2024},
  publisher={IEEE}
}

@article{fu2024mogo,
  title={Mogo: Rq hierarchical causal transformer for high-quality 3d human motion generation},
  author={Fu, Dongjie},
  journal={arXiv preprint arXiv:2412.07797},
  year={2024}
}

@inproceedings{fan2022faceformer,
  title={Faceformer: Speech-driven 3d facial animation with transformers},
  author={Fan, Yingruo and Lin, Zhaojiang and Saito, Jun and Wang, Wenping and Komura, Taku},
  booktitle={Proceedings of the IEEE/CVF conference on computer vision and pattern recognition},
  pages={18770--18780},
  year={2022}
}

@inproceedings{cudeiro2019capture,
  title={Capture, learning, and synthesis of 3D speaking styles},
  author={Cudeiro, Daniel and Bolkart, Timo and Laidlaw, Cassidy and Ranjan, Anurag and Black, Michael J},
  booktitle={Proceedings of the IEEE/CVF conference on computer vision and pattern recognition},
  pages={10101--10111},
  year={2019}
}

@article{zhou2025hand,
  title={Hand Gesture Recognition From an Open-Set Perspective},
  author={Zhou, Jun and Xu, Chi and Cheng, Li},
  journal={IEEE Transactions on Multimedia},
  volume={27},
  pages={4181--4192},
  year={2025},
  publisher={IEEE}
}

\vfill

\end{document}